\documentstyle[floats,tighten,prl,aps,epsf]{revtex}
\draft
\begin{document}

\twocolumn[\hsize\textwidth\columnwidth\hsize\csname
           @twocolumnfalse\endcsname

\def\Zsun{\thinspace\hbox{$\hbox{Z}_{\odot}$}}
\def\msun{\thinspace\hbox{$\hbox{M}_{\odot}$}}
\def\rsun{\thinspace\hbox{$\hbox{R}_{\odot}$}}
\def\lsun{\thinspace\hbox{$\hbox{L}_{\odot}$}} \def\.{\'{\i}}

\title{Gravitational-Wave Bursts Induced by Neutrino Oscillations: The
Origin of Asymmetry in Supernovae Explosions}

\author{Herman J. Mosquera Cuesta}

\address{{High Energy/Astrophysics Section\\ Abdus Salam International
Centre for Theoretical Physics  \\ Strada Costiera 11, Miramare 34014,
Trieste, Italy \\ e-mail: herman@ictp.trieste.it}}

\date{\today}

\def\be{\begin{equation}} \def\ee{\end{equation}} \maketitle

\begin{abstract}
\widetext  If neutrino   flavor   changes   really  exist,  to    say:
$\mu$-neutrino oscillating into a {\it sterile} neutrino, then, it can
be expected  that   due to  neutrino oscillations  and   non-spherical
distortion  of the resonance  surface  induced by the magnetic  field,
some  asymmetric emission of sterile  neutrinos  can occur during  the
protoneutron star formation at the onset of a supernova core-collapse.
Assuming no strong suppression  of the oscillations, the non-spherical
huge neutrino  energies released, ($\sim  10^{53-54}\; erg$), together
with the   proto-neutron  star rapid  rotation,   may trigger powerful
bursts of gravitational waves  by the time neutrino flavor conversions
ensue.   I show  here  that these bursts    are detectable by the  new
generation  of gravitational-wave  detectors as LIGO,  VIRGO and TIGAs
for  distance scales  $\sim 10\; kpc$.    It  is also  argued that the
general relativity requirement of  an ellipsoidal axisymmetric core at
maximum gravitational-wave  emission  induced by  $\nu$-luminosity can
properly   be afforded   by  the  neutrino-sphere   geometry  when the
oscillations onset.  The connection of  neutrino oscillations with the
supernova   asymmetry and bi-polar jets   ejecta is shown naturally to
appear in this scenario.
\end{abstract}

 \vskip 2pc
\narrowtext
]

\def\be{\begin{equation}} \def\ee{\end{equation}}

Turning  the millenium   a   new generation  of    interferometric and
resonant-mass gravitational-wave antennas   will be operational.    If
confirmed, the detection     of gravitational radiation   will make  a
breakthrough             in          contemporary   astrophysics   and
cosmology\cite{thorne95,schutz95}. Due to the fundamental impact these
observations would have, investigation of potential detectable sources
becomes a key piece.  In this {\it  letter} I propound for first  time
that  a new  fundamental source of   gravitational radiation  had been
overlooked by  the community of workers on  the fields. I suggest that
the recently   confirmed  resonant conversion  of neutrinos  in  dense
magnetized matter  can be a very important  new one. I conjecture that
this   neutrino temperature   asymmetry,   produced  during   $\nu_\mu
\leftrightarrow    \nu_{\tau,s}$ flavor    changes,   couples  to    a
non-spherical mass-energy    distribution (the  oscillating   neutrino
fluid) inside the  proto-neutron star during a core-collapse supernova
explosion. The resonant  neutrino  conversion and rapid   rotation may
provide    the  key ingredient  for  the   analogous  of a quadrupolar
mass-moment tensor for the  whole   NS to  develop, as  currently   is
pictured    in modeling  GWs   from nonaxisymmetric   spinning neutron
stars.  Consequently,    it makes   it  possible  the   generation  of
gravitational-wave bursts  induced  by neutrino conversions,  $\nu_\mu
\leftrightarrow \nu_{\tau,s}$, powered by the onset of the collapse of
the inner core during explosions of massive stars.

In  this   {\it letter} it   is argued  that the  neutrino oscillation
process as described by Kusenko and Segr\`e\cite{segre98} (see below),
and its associated   gravitational-wave  burst, as suggested  for  the
first time here, occurs  during the very  early stages of evolution of
the proto-neutron star (PNS), $\leq 10\; ms$ after  the core bounce at
$\rho  >  2  \times  10^{14}\;  g  cm^{-3}$   (see Burrows  and  Hayes
\cite{burrows95} and Epstein\cite{epstein78}, and references therein).
The  evolution of  the  rapidly rotating   PNS  during this  phase  is
dominated by  vigorous entropy-driven convective motions.  Contrary to
the arguments given  by Janka and Raffelt \cite{raffelt99}, concerning
the existence of a structured PNS  (mantle and atmosphere), we suggest
that at this post-bounce evolutionary stage there  is no room for such
a stratified   structure  to appear.    Simply,  convection  (overturn
timescale $\sim 1\; ms$) dominates the  PNS evolution, what results is
a  nonx  structured  PNS  at   all.    Consequently, the  analysis  of
Ref.\cite{raffelt99} is not    applicable to our model.   Instead,  we
follow   the lines  of   Kusenko and  Segr\`e  \cite{segre98}  in what
concerns to  the  oscillation mechanism itself and   its implications.
Furtherly, since  we expect the  neutrino oscillation-driven GWs burst
to be emitted  at core  bounce (mass $  M_{core}  =  1.4 \;  M_\odot$,
radius $R \sim 10^7  \; cm$\cite{wilson76}, timescale $\tau_B \sim 1\;
ms$\cite{turner77}), its timescale is a very short one compared to the
neutrino diffusion  time (the PNS  Kelvin-Helmholtz neutrino  cooling)
$\sim   10\; s$,  the  process  itself  can be  considered   as a {\it
transient    stage}\cite{raffelt99}.         As  pointed    out     by
Ref.\cite{raffelt99} for this  phase the mechanism proposed by Kusenko
and Segr\`e\cite{kusenko97,segre98} may properly be at work.

The issue of neutrino oscillations has quite  recently been invoked to
explain  a variety of astrophysical  and cosmological concerns. Recent
researches  in the  astrophysics of  compact objects  allow Kusenko \&
Segr\`e\cite{segre98}    and    Grasso,     Nunokawa    \&       Valle
(GNV98)\cite{valle98} to  claim   that the pulsar velocities   (recoil
kicks) may be explained assuming anisotropic neutrino emission induced
by  $\nu_{\mu,\tau}$   oscillations      into $\nu_e$,      and   also
sterile-to-active oscillations of  massive\cite{kusenko97,segre98}  or
massless\cite{valle98} neutrinos, provided a strong magnetic field ($B
\sim  10^{15}$    G) be  involved.   Meanwhile,  Akhmedov,   Lanza and
Sciama\cite{valle98}  have   studied the influence    of resonant spin
flavor  precession   of  neutrinos   during    core-collapse supernova
explosion leaving  neutron stars as remnants. Their   idea is based on
the Mikheev, Smirnov \& Wolfenstein (MSW) neutrino resonance\cite{MSW}
occurring on a surface off-centric to a proto-neutron  star core.  The
neutrino      relative        recoil     momentum     is       defined
as\cite{kusenko97,segre98}

\be 
\frac{\Delta \kappa}{\kappa} = \left[ \frac{  4 e \sqrt{2}}{\pi^2}\right]
\left(\frac{\mu_e \mu{^{1/2}_n}}{m{^{3/2}_n}  T^2} \right)  B  =  0.01
\left(\frac{B}{1.2 \times 10^{15}{\rm G}}\right).  
\ee

In this equation $\mu_e$ and $\mu_n$, define  the electron and neutron
chemical potential, respectively,  $m_n$ is the  neutron mass, and  $T
\sim 30$   MeV  the  temperature  at   the   $s$-neutrinosphere.  This
anisotropic   temperature distribution  (the   matter-induced neutrino
potentials) is created by the strong magnetic  field ($B \sim 10^{15}$
G)\cite{kusenko97} developed during the  supernova core collapse.  So,
the resonant   conversions  of ordinary  neutrinos  ($\nu_\mu$)   into
sterile ones ($\nu_s$),  if they exist,  can occur all the way through
the  proto-neutron  star (PNS)  structure   involving huge  amounts of
non-spherically  distributed energy.  This asymmetric sterile neutrino
energy configuration is a   crucial point for the arguments  presented
below  supporting the idea that   some gravitational-wave (GWs) bursts
are expected to be  generated by the  same time  the active-to-sterile
neutrino oscillations develop.   This $\nu$-oscillation  view seems to
be confirmed by  the Super-Kamiokande Neutrino Experiment (SNE), whose
most recent results, published  by Fukuda et  al.\cite{fukuda98}, have
presented compelling evidence for neutrino oscillations.

Core-collapse supernovae  explosions  are one   of the  most  powerful
sources of neutrinos $\nu_e,  \nu_\mu, \nu_\tau$ and  probably $\nu_s$
and its antiparticles.  Different  theoretical and numerical models of
type II  supernovae   explosions \cite{woosley}  have  estimated  that
$\Delta  E_{total} = 5.2 \times  10^{53}  erg (\frac{10\; km}{R_{NS}})
(\frac{M_{NS}}{ 1.4 \; M_\odot})^2$ are carried away by neutrinos (see
below). Almost $\sim  10^{58}$  neutrinos  of mean energies   $(10-25)
\;MeV$  are released over time scales  of seconds. Investigations have
shown that nearly 99\% of the total  gravitational binding energy of a
protoneutron star can   directly be carried  away  by these  neutrinos
during the supernova explosion\cite{woosley,janka96}, the  remaindings
streaming  away as photons  and  gravitational waves \cite{janka96} -a
small  part of  it  $\sim  (10^{-6}-10^{-7})  \; M_\odot$-   generated
through the  quadrupolar  distortion of  its  mass tensor.  Because of
$\nu_\mu$ and  $\nu_\tau$  interact only  via neutral   currents their
neutrino-spheres are deeper in the core, therefore their mean energies
are higher,  near (15-25) MeV.    At this  energy level, detection  of
$\nu_e$  and $\bar{\nu}_e$  may   occur since they basically   undergo
charged-current  interactions, while  as  quoted  above $\nu_\mu$  and
$\nu_\tau$  interact through  their  neutral-currents\cite{MSW}.  This
last point reinforces why their detection  is crucial to test the {\it
supernova mechanism}.

In nuclear matter  optical depths (mean  free paths) are different for
different    neutrino species. Then   they  can     undergo  $\nu_e
\longleftrightarrow     \nu_{\mu,\tau}$     conversions    through the
charged-current interactions. Whenever mixing  and $\nu_\tau$ mass are
non-vanishing, the spherical   symmetry of the resonance surface   for
$\nu_e \longleftrightarrow \nu_\tau$  oscillations there doesn't exist
any  longer, and an asymmetric  distribution  of the emitted  neutrino
momenta can occur. The  resonance condition for conversions in between
neutrinos of left-handed of flavor $i$ into right-handed of flavor $j$
($i,j$ = $e$, $\mu$, $\tau$ or  $s$, with $s$ representing the sterile
neutrino) is given by\cite{MSW,smirnov99}

\be  V(\nu_{iL}) + \frac  {m{^2_{v_i}}}{2 E_\nu} = V(\bar{\nu}_{jR}) +
\frac {m{^2_{v_j}}}{2 E_\nu},  \ee

with $V(\nu_s) = 0$. In the case of magnetized material media it reads:

\be  \sqrt{2}G_F N_{eff} -  c_{eff} B_{||} = \frac{\Delta m^2}{2E_\nu}
\cos{2\theta_0}.  \ee

 An extensive discussion  of neutrino physics including mean potential
energies of active neutrinos  and  antineutrinos is approached  to  in
Ref.\cite{smirnov99}, and  references  therein.  In the  last equation
the effective  parameters for the  resonant  conditions, $N_{eff}$ and
$c_{eff}$,  depend   on  the     exact   nature of    the  oscillation
involved\cite{kusenko97}.   Since  the mass difference constraints the
occurrence of a specific resonant conversion, we assume a hierarchical
neutrino  mass scheme: $ \nu_s  >>  \nu_\tau >> \nu_\mu$.  Thus $\nu_s
\leftrightarrow   \bar{  \nu}_{\tau,\mu}$  neutrino  oscillations  are
allowed.

However,  Smirnov  \cite{smirnov99} has   called the attention  on the
possibility   that   a     very     strong   suppression    of     the
$\nu_{\mu}-\nu_{\tau,s}$ mixing angle can occur in this picture due to
the fact that  for transitions in extremely dense  matter  such as the
ones prevalent  in  the supernova  at   core bounce,  $\rho_c  \approx
(10^{14}-10^{15}) g/cm^{3}$, a rather low mixing angle develops, since
the adiabaticity condition should be fulfilled for oscillation lengths

\be L_{\nu_\mu \rightarrow  \nu_s} \simeq \left([{(2\pi)}^{-1}\frac  {
\Delta   m^2}{  2\kappa}]    \sin{2\theta}\right)^{-1}         \approx
\left(\frac{10^{-2}}{\sin{2 \theta}}\right) cm, \ee

 smaller than  the  typical density  variation  scale-height ($H_{N_e}
 \sim    6\;$ km).   This  constraint is    kept if  the  mixing angle
 $\sin^2 {2\theta} >   10^{-8}$.  It  is easy to    check  the degree  of
 suppression to be  around eight orders of  magnitude smaller than the
 one  Super-Kamiokande  Experiment has   inferred to.  However,  since
 during its ploughing through  the PNS core $\nu_\mu$-neutrinos can be
 absorbed and reemitted $\sim 10^5$  times, the effective  suppression
 of the $\nu_{\mu}-\nu_{s}$  mixing  angle reduces  to $\sim  10^{-3}$
 compare to SKE.   This is consistent with  the upper limit derived in
 Ref.\cite{kainulainen91} for the mass range  relevant to our problem:
 $\Delta m \sim  (3-10)$ keV, for which the  mixing angle turns out to
 be:  $\sin^2 {2\theta}_{\bar{e},s}   >  10^{-6}$.  For  larger mixing
 angle, neutrino  detections   from  SN1987A by  IBM    and Kamiokande
 experiments   would    render  contradicted   since    the   electron
 anti-neutrinos get  reduced because most  of the  PNS energy would be
 carry    off      by   the        sterile   neutrino.     In      the
 $\bar{\nu}_{e}\leftrightarrow \nu_s$ case, it is expected that no
 more than  one   third  of  the  total    neutrino flux is    emitted
 asymmetrically\cite{kusenko97}.   Moreover,   because  the  explosion
 produces  $\sim 10^{58}$ neutrinos,   the minute parcel available can
 still  change flavor forwards  and reverse back  several times when a
 radial  distance from the  proto-neutron star  edge equivalent to one
 mean free  path  has been   reached.   There onwards  the oscillating
 sterile neutrinos will   carry out information concerning the  highly
 distorted neutrino-sphere  at  the point  of their  last  conversion.
 This    is the    source     of  the    neutrino   oscillation-driven
 gravitational-wave burst.

Simulations   of supernovae explosions     have demonstrated that  the
geometry of  the   neutrino-sphere is not  the   one of a    {\it real
sphere}\cite{janka96}.  High density gradients  in the compact  object
produced as the implosion goes on create anisotropic mass distribution
regions where the matter  density surpasses the required  for neutrino
trapping (see\cite{klapdor,woosley}),  while   in others  they  can go
throu freely.   That time-varying anisotropic  distribution of density
gradients  suggest that  the energy-momentum   tensor of the  neutrino
fluid is somehow quadrupolar. Then, its time variation translates into
the  equivalent  of   a changing  quadrupole  tensor   of  mass, whose
propagation induces emission of gravitational   waves\cite{burrows95}.
The other  way round, the  supernova magnetic field induces the matter
content polarization what forces resonant conversions to take place at
very radial distances from the star core  for those neutrinos released
along the  magnetic field axis. This  yields in a rather inhomogeneous
distribution of temperatures for the resonant surface over the dipolar
axis, what induces an  asymmetric spatial distribution of the radiated
neutrino momenta, and with  this the possibility that  a GWs burst can
be triggered.

For the  above scenario, we  can estimate  the  characteristics of the
gravitational radiation  emitted  by using  Newtonian  gravity and the
quadrupole approximation to  general relativity. Following Burrows and
Hayes\cite{burrows95} get get

\be   h{^{tt}_{ij}}  = \frac{4  G}{c^4  D} \int^t_{-\infty} \alpha(t')
L_\nu(t') dt'.  \ee

Here $h{^{tt}_{ij}}$ represents the transverse-traceless dimensionless
metric strain,  $D$  defines the   distance to  the source, $0.2  \leq
\alpha(t)  \leq 0.8$\cite{burrows95}  is the instantaneous  quadrupole
anisotropy, $L_\nu(t)$  the overall luminosity  of neutrinos.   In our
model, we associate the source of the  gravitational-wave burst to the
quadrupolar moment of the  neutrino energy flow, $\Delta  E_{\nu}$, by
the time the oscillation occurs inside the proto-neutron star, instead
of to the PNS quadrupole  tensor of mass, as  usually done in ordinary
neutrino-driven supernova explosion models\cite{janka96}. With     the
assumption   that the  neutrino   oscillation timescale is $T_{\nu_\mu
\longleftrightarrow \nu_s} \sim c/R_{PNS}  \sim 1.0\times 10^{-3} \;s$
at the  PNS interior, where $R_{PNS}$ is  the PNS radius,  then we can
expect the  burst  waveform of  the radiation emitted  to look  like a
delta Dirac function centred around 1 kHz, superimposed to the overall
waveform  as computed by Burrows  and Hayes\cite{burrows95}.  Then the
characteristic, normalized, gravitational-wave    amplitude  generated
during the supernova  explosion   due to the  magnetically   distorted
(non-spherical) outcoming front of  the $s$ neutrino-sphere is $h \sim
1.4  \times 10^{-23}\; {\rm Hz}^{-1/2}$,   for a source distance $10\;
kpc$, and in  the   case for which   100\% of  the neutrinos   undergo
oscillations  in  between  $\nu_\mu \longleftrightarrow \nu_{\tau,s}$,
with  luminosity  $L_\nu = 3\times   10^{54}\; erg/s$   released on  a
timescale    $3   \times   10^{-2}\;   s$  \cite{epstein78,burrows95}.
Unfortunately, a  gravitational-wave signal  from the Large Magellanic
Cloud would not be detectable.

Therefore, a gravitational-wave signal such as this will be detectable
by the new generation of   gravitational-wave detectors such as  LIGO,
VIRGO and TIGAs  detectors\cite{thorne95,schutz95} provided the  large
part  of neutrino undergo oscillations on  a  timescale $\sim 1 \; ms$
(see  Ref.\cite{turner77}). The gravitational   wave frequency  of the
main signal (GW burst) produced will be $f_{gw} \sim 1 $ kHz.

If   the  mechanism   neutrino oscillations/gravitational-wave  bursts
really  realizes in nature,  then, it  can  afford a natural bridge to
link and explain  two,  until  now, difficult problems  in  supernovae
physics,    i.  e., a)  what  is  the  very origin  of the asymmetries
observed  in supernovae ejecta and  remnants?  b) what did trigger the
bi-polar   beamed       jets    demonstrated    by     Nisenson    and
Papaliolios\cite{nisenson99} to be in association  with SN 1987A? (see
Ref.\cite{cen99}).  It is  notheworthy that similar bi-polar  jet-like
structures have been observed in many  other supernovae remnants.  For
a complete account  of this  issue see Gaensler\cite{gaensler99},  and
references  therein).  The   general relativistic   description of the
gravitational   radiation generated  by  escaping   neutrinos during a
supernova    explosion,  as  demonstrated  by Epstein\cite{epstein78},
requires  that  the collapsing  star be ellipsoidally  deformed due to
rotation  and anisotropic  $\nu$-emission.    In this case  the metric
strain can be expressed by\cite{epstein78}

\be    |\theta^{ij(\nu)}_{TT}|  \sim     3.6   \times   10^{-21}   e^2
\left(\frac{55\; {\rm kpc}}{D} \right).\ee

where $e$ corresponds  to the eccentricity  of the ellipsoid projected
in    the  plane through which   the   oscillating  $\nu$s escape.  We
conjecture here that such an asymmetry $e$ is determined fundamentally
by  the  shape of  the  neutrino-sphere by   the time  the oscillation
develops.   The other  way round,   the anisotropic neutrino  emission
responsible, in the picture by  Kusenko and Segr\`e\cite{segre98}, for
the nascent  neutron star kicks,  might be considered the appropriated
scaling  of the eccentricity   appearing  in eq.(5)  during supernovae
explosions.     If it    is  the   case,  using   Kusenko  and Segr\`e
\cite{kusenko97,segre98} results for $B = 10^{15}\; G$, we can write

\be e^2 \sim \frac{\Delta \kappa}{\kappa}= 0.01, \ee

from which we derive the angle $\theta$  the final jets will be beamed
into driven by neutrino oscillations

\be \tan   \theta \simeq e  =  \frac{a-b}{a+b}  = 0.1  \longrightarrow
\theta = 5.7 \; deg.  \ee

This  interesting  physical connection   might  provide a  simple  and
elegant   explanation  to    the    points   raised   by   Wang    and
Wheeler\cite{ww98} concerning the  observational features  of SN1998bw
and   other SNe.  Based upon   spectropolarimetric  studies of a large
sample of  SNe, these authors  suggested that sufficiently high energy
deposition  ($\sim  10^{50}\;  erg$)  along the polar   axis can build
oblate isodensity contours  in central  core  collapses, and that  the
formation    of  such  structures   may    lead  to  produce  jet-like
structures. In our view,   the  overall process  will  follow  as: the
stalled     supernova   explosion   is  revived   by   neutrino-driven
convection\cite{janka96}. Some instants before, the flavor conversions
occur,    and  the  large    part   of  $\nu$s   is  absorbed  through
$\beta$-interactions\cite{segre98}, transferring   that way  the  huge
momenta and energy  they carry out.  The baryon  matter in the  mantle
absorbing the  oscillating  $\nu$s will keep,  and transport outwards,
the memory of  the asymmetric neutrino  absorption.   Thus, when these
layers finally drive the SNe explosions themselves, they would reflect
the aspherical momenta gained at the core.  Since, as shown above, the
large $\nu$  flux  is strongly collimated, it   is likely then  that a
strong jetted matter ejecta, and a highly polarized expanding luminous
envelope appears. An additional bonus given by the present scenario is
the fact the model naturally explains why do the expected emitted jets
are to  have markedly different  strengths, a point  also discussed by
Cen\cite{cen99} in his unified  picture  for pulsar, gamma-ray  bursts
and supernovae. In our picture, this power asymmetry of jets is fundamentally created by the neutrino oscillation mechanism.

To conclude, we  have  propound for the   first  time that a  very
fundamental new class of   gravitational wave bursts can  be  produced
during   resonant {\it  ordinary-to-sterile}    neutrino  oscillations
($\nu_\mu \leftrightarrow \nu_{\tau,s}$) inside proto-neutron stars at
the onset  of core-collapse  explosions  of massive stars.    This new
family of  GWs signals should  be viewed  as one  of a  very different
nature  when   compared  to   the   classical  quadrupolar mass-tensor
gravitational waves,   as currently is  suggested  in  studies on this
subject,   when  ordinary    neutrinos   drive  the  final   (stalled)
explosion. The characteristic properties of the signal are found to be
in the sensitivity  range of most of  the now under construction or in
phase of planning gravitational-wave observatories.

This  mechanism  for  GWs   emission was  not  envisaged    in  former
investigations on the subject  perhaps because there was no compelling
evidence  for the neutrino flavor changes.   This letter introduces it
for  the  first time,  in  the  context  of mechanisms  for generating
gravitational   radiation   from astrophysical sources.   The eventual
detection of such bursts in future experiments  will be a breakthrough
in  fundamental physics  for, as  pointed  out above,  there exist the
chance  that an important parcel  of the oscillating sterile neutrinos
can  get  the Earth nearly  by  the time  the associated gravitational
radiation  bursts  they  produced  when   changing flavors
$(\nu_\mu \leftrightarrow \nu_{\tau,s})$ arrive  at the detectors. 

\acknowledgements{I am    grateful to Dr.    A.  Kusenko and Professor
A. Yu.  Smirnov for their  kindly  reading  of this  manuscript and 
valuable key  suggestions.}


\begin{thebibliography}{99}

\bibitem{thorne95}  K.  S.  Thorne, {\it   Proceedings of the Snowmass
Summer Study on Particle and Nuclear Astrophysics and Cosmology}, eds.
E.  W.  kolb and R.   Peccei, Wprld Scientific, Singapore (1995). {\it
ibid} {\it Three Hundred Years of  Gravitation}, Eds.  S.  W.  Hawking
and W.  Israel, Cambridge University Press, Cambridge, England (1987).

\bibitem {schutz95}B.F. Schutz, in {\it Proceedings of the Les Houches
School  on   Astrophysical Sources of Gravitational   Radiation}, eds.
J.-P. Lasota and  J.-A.  Marck, Cambridge University Press, Cambridge,
England   (1995). B.F.    Schutz,  report astro-ph/9802020,   February
(1998).


\bibitem {segre98}A. Kusenko  and G. Segr\`e,  Phys.  Rev. Lett.  {\bf
79}, 2751 (1997).  {\it ibid.}   {\bf 77}, 4872  (1996), Phys. Rev. D,
{\bf   59},  061302 (1999).  



\bibitem{burrows95}  A. Burrows  and J.  Hayes,  Phys. Rev. Lett. {\bf
76}, 352 (1996).   See also A. Burrows, J.  Hayes and B.  A.  Fryxell,
Astrophys. J. {\bf 445}, 830 (1995).

\bibitem{epstein78} R. Epstein, Astrophys. J., {\bf 223}, 1037 (1978).

\bibitem{raffelt99} H.Th. Janka and G. G.  Raffelt, Phys. Rev. D, {\bf
59}, 023005 (1999).

\bibitem{wilson76}J. R. Wilson, Astrophys. J., GREP Conference 79 (1977).

\bibitem{turner77}M. Turner and R. V. Wagoner, Stanford University IT,  preprint No. 576 (1977).



\bibitem{kusenko97} A. Kusenko  and   G. Segr\`e, Phys.  Letts.  B{\bf
396}, 197 (1997).

\bibitem{valle98} D.   Grasso,  H.  Nunokawa  and J.W.F. Valle,  Phys.
Rev.  Lett. {\bf 81}, 2412 (1998).  See also  E.Kh. Akhmedov, A. Lanza
and D.W. Sciama,  Phys.  Rev. D {\bf  56}, 6117 (1997).  H.  Nunokawa,
V.B.    Semikoz,  A.Yu.  Smirnov  and  J.W.F.  Valle, Nucl. Phys. {\bf
B501}, 17 (1997).

\bibitem  {MSW}S.P. Mikheyev and  A.Yu.  Smirnov, Yad. Fiz.  {\bf 42},
1441 (1985).  [Sov.   J.  Nucl.  Phys.  {\bf  42}, 913 (1985)];  Prog.
Part.  Nucl.   Phys.     {\bf  23},   41   (1989).   L.   Wolfenstein,
Phys. Rev. D. {\bf 17}, 2369 (1978); {\bf 20}, 2634 (1979).


\bibitem {fukuda98}Fukuda, Y., et al., Phys. Rev. Lett. {\bf 81}, 1562
(1998).


\bibitem {woosley}S.E. Woosley and T.A. Weaver, Astrophys. J. S., {\bf
101}, 181 (1995). A. Burrows, AAS, {\bf 187}, 1704 (1995). {\it ibid.}
{\bf 187}, 1703 (1995).

\bibitem {janka96}H.-Th.  Janka and E.  M\"{u}ller, A  \& A {\bf 306},
167 (1996). E.   M\"{u}ller and H.-Th.  Janka, A  \& A {\bf  317}, 140
(1997).


\bibitem  {smirnov99}Private   communication  with Professor   A.  Yu.
Smirnov, International Centre for   Theoretical Physics  (ICTP).   See
also J.F.  Beacom, {\it Supernova  Neutrinos and the Neutrino Masses},
report hep-ph/9901300, January (1999).

\bibitem{kainulainen91}K. Kainulainen, J. Maalampi and J. T. Peltoniemi, Nuc. Phys. B, {\bf 358}, 435 (1991).

\bibitem  {klapdor}H.V.   Klapdor-Kleingrothaus  and K.   Zuber,  {\it
Particle Astrophysics},  Chapter  13, Institute of Physics  Publishing
Ltd., Bristol and Philadelfia, USA (1997).


\bibitem{nisenson99}   P.   Nisenson  and   C.    Papaliolios,  Report
Astro-ph/9904109, April 9 (1999).

\bibitem{cen99} R.   Cen, Astrophys. J.,  {\bf   507}, L131 (1998) and
Report Astro-ph/9904147, April 12 (1999).

\bibitem{gaensler99}B. M.   Gaensler,   Ph.  D.   Thesis Dissertation,
University of Sidney, Australia. Unpublished (1999).

\bibitem{ww98} L. Wang and J.  C.   Wheeler, Astrophys. J., {\bf 504},
L87 (1998). See also J.C.  Wheeler, P.  Hoeflich and L.  Wang, AAS DPS
meeting \# 193, \# 42.06 (1998), Report Astro-ph/9904047, April 3 (1999).



\end{thebibliography}
\end{document}